# Size, nanostructure and composition dependence of bimetallic Au-Pd supported on ceria-zirconia mixed oxide catalysts for selective oxidation of benzyl alcohol

*Carol M. Olmos[a], Lidia E. Chinchilla[a], Alberto Villa[b], Juan J. Delgado[a], Ana B. Hungría[a], Ginesa Blanco[a], Laura Prati[b], Jose J. Calvino[a], Xiaowei Chen[a,*]*

*a Departamento de Ciencia de los Materiales, Ingeniería Metalúrgica y Química Inorgánica, Facultad de Ciencias, Universidad de Cádiz, Campus Río San Pedro, Puerto Real (Cádiz), E-11510, Spain*

*b Dipartimento di Chimica, Università degli Studi di Milano, Milan, I-20133, Italy*

[*] Correspondence to: Dr. X. Chen, tel: 0034-956-012741, fax: 0034-956-016288, Email: xiaowei.chen@uca.es.

**Abstract**

A bimetallic Au-Pd catalyst supported on ceria-zirconia with Au:Pd molar ratio of 0.8 has been synthesized using simultaneous deposition-precipitation method and oxidized at 250, 450 and 700 °C in order to modify its particle size, nanostructure and composition. The combined X-ray Energy Dispersive Spectroscopy (XEDS) and X-ray photoelectron spectroscopy (XPS) analysis clearly evidence that the bimetallic Au-Pd catalyst oxidized at 250 °C is made up of a mixture of monometallic Au and Pd and bimetallic Au-Pd nanoparticles with Au:Pd ratios varying in a wide range. Increasing oxidation temperature leads to a larger interaction between Au and Pd. Meanwhile, a slight increase of particle size and a narrowing of the Au:Pd ratio in the bimetallic nanoparticles take place. Compared with titania and activated carbon supports, the resistance against sintering at high temperatures of Au-Pd metal particles supported on ceria-zirconia is proven to be higher. A synergistic effect has been observed for selective oxidation of benzyl alcohol on these catalysts. The catalytic activity decreases only slightly after oxidation at 450 °C. However, oxidation at 700 ºC results in much lower catalytic activity. Migration of Pd onto Au particles during oxidation of benzyl alcohol enhances the catalytic activity of a physical mixture of monometallic Au and Pd supported on ceria-zirconia catalysts. This fact, jointly with an analysis of the intrinsic activity, reveals an influence of the actual nature of Au-Pd interactions in the bimetallic particles, which points out to a higher activity of Au@Pd or AuPd@Pd nanostructures on ceria-zirconia support.

# 1. Introduction

Selective oxidation of benzyl alcohol to benzaldehyde in liquid phase using heterogeneous catalysts has shown many advantages, such as easy recovery and reusability of catalysts and high catalytic activity [1]. Among these catalysts, Au-Pd nanoparticles supported on different kinds of supports have exhibited very attractive results [2–11].

Even though plenty of articles related to bimetallic Au-Pd catalysts for benzyl alcohol oxidation have been published, only a few of them focus on reducible ceria-zirconia mixed oxide supports [10,11]. Our previous work reported the occurrence of a synergistic effect between Au and Pd supported on ceria-zirconia, resulting in a catalyst with higher activity than other commonly used supports, such as activated carbon, carbon nanotubes and titania for selective oxidation of benzyl alcohol [11]. For thermal treatments at 250 ºC, oxidizing or inert gas leads to a catalytic activity higher than that of the fresh catalyst reduced by $NaBH_4$ or that of the catalyst reduced in hydrogen. However, the frequency of bimetallic Au-Pd particles in these catalysts accounts for only 25-40%, which is very likely due to the limited solubility between Au and Pd [11].

Very recently, Carter et al. reported the synergistic effect between Au and Pd on a similar support with different Au:Pd molar ratios for the same reaction [10]. The highest catalytic activity was achieved on a Au-Pd/$CeZrO_4$ catalyst with total metal loadings of 2wt.% and Au:Pd molar ratio of 44:56 [10]. They also stated that Pd monometallic nanoparticles were much harder to be visualized due to its poor mass contrast against heavy $CeZrO_4$ support, as we commented in our previous articles [11–13]. The actual complexity of bimetallic Au-Pd system on different supports, in particular on ceria-zirconia mixed oxide support, both in terms of metal particle composition, size and morphology, has also been pointed out in the literature [7-9,11,12]. Actually, the mixing patterns of Au and Pd can be core-shell, subcluster segregated alloy, ordered or random alloy [15]. It is therefore clear that a statistical analysis, including a large number of metal particles, is mandatory to stablish any correlation with performance and to determine the actual nanostructure of this type of complex catalysts.

An approach which takes into consideration this complexity is still lacking in the analysis of the activity of bimetallic Au-Pd catalysts on ceria-zirconia mixed oxide support for benzyl alcohol oxidation. In addition, the influence of the size, composition and morphology of Au-Pd nanoparticles on the catalytic activity for benzyl alcohol oxidation has not been studied in details.

In this work, the size and homogeneity of bimetallic Au-Pd particles have been tuned by increasing oxidation temperature a bimetallic Au-Pd catalyst supported on a $Ce_{0.62}Zr_{0.38}O_2$ mixed oxide with a Au:Pd molar ratio of 0.8. A mostly composed of bimetallic, alloy-type, nanoparticles with composition distributed homogeneously has been successfully synthesized. Monometallic Au and Pd catalysts and a physical mixture of Au and Pd on ceria-zirconia catalyst was used as reference samples. The correlation between metal particle size, metal particle composition and catalytic activity is discussed on the basis of an in-depth characterization work which considers a very large sets of analyzed nanoparticles. To the best of our knowledge, no such systematic and in-depth study on size, composition and nanostructure of bimetallic Au-Pd supported on a reducible ceria-zirconia mixed oxide catalysts for benzyl alcohol oxidation has been performed.

## 2. Experimental

*2.1. Catalyst preparation*

The precursors, $HAuCl_4·3H_2O$ (99.99 %) and $PdCl_2$ (99.99 %), were purchased from Alfa-Aesar and Sigma-Aldrich, respectively. The $Ce_{0.62}Zr_{0.38}O_2$ support was kindly provided by Grace Davison. The preparation method of the reference monometallic Au and Pd supported on ceria-zirconia has been described in our previous work [12,13]. The monometallic 2.5 wt.% $Au/Ce_{0.62}Zr_{0.38}O_2$ (Au/CZ) catalyst was prepared by deposition-precipitation method with $Na_2CO_3$ as precipitating agent [12,13]. The monometallic 1.2 wt.% $Pd/Ce_{0.62}Zr_{0.38}O_2$ (Pd/CZ) catalyst was prepared by incipient wetness impregnation using an aqueous solution of $Pd(NO_3)_2•4NH_3$. The impregnated sample was dried at 100 ºC overnight, calcined at 350 ºC for 4 h in a muffle oven and finally reduced at 350 ºC under flowing 5% $H_2/Ar$ for 1 h.

The bimetallic Au-Pd catalyst with nominal Au:Pd ratio of 1:1 was prepared using simultaneous deposition-precipitation method [12]. $PdCl_2$ was dissolved in a 1 M HCl solution and then mixed with the aqueous solution of $HAuCl_4$. The $HAuCl_4$ concentration was 3.2 mM. Using a liquid pump, the mixture of the $HAuCl_4$ and $PdCl_2$ solutions was added to a flask containing 15 g of the $Ce_{0.62}Zr_{0.38}O_2$ support suspended in 800 mL deionized water. An automatic titrator TIM 856 was used to adjust and maintain the pH value at 8 with an aqueous 0.05 M $Na_2CO_3$ solution. The sample was subsequently dried at 100 ºC overnight. The dried catalyst was oxidized in a flow of 5% $O_2/He$ for 1 h at 250 ºC. Then the gas flow was switched to He and kept at the same temperature for 1 h. This $Au-Pd/Ce_{0.62}Zr_{0.38}O_2$ catalysts with actual Au:Pd molar ratio of 0.8 is referred as to 0.8AuPdCZO250. The bimetallic 0.8AuPdCZO250 catalyst was also oxidized in a flow of 5% $O_2/He$ at 450 and 700 ºC for 1 h. O450 and O700 were added to the name of these catalysts to indicate the oxidation temperature. The actual Au and Pd loadings and Au:Pd ratios of the different catalysts are listed in Table 1.

*2.2. Catalyst characterization*

Textural properties of all the catalysts were determined by $N_2$ adsorption at -196 °C using Micromeritics ASAP2020. Inductively coupled plasma-atomic emission spectrometry (ICP-AES) was employed to measure the Au and Pd loadings of the catalysts. X-ray diffraction (XRD) analyses of the catalysts were carried out using a Bruker diffractometer Model D8 ADVANCE operated at 40 kV and 40 mA employing Cu Kα radiation.

XPS analyses were performed on a Kratos Axis Ultra DLD instrument. Spectra were recorded using monochromatized Al Kα radiation (1486.6 eV), with an X-ray power of 150 W. The spectrometer was operated in the Constant Analyzer Energy mode, with a pass energy of 20 eV. Powder samples were pressed into pellets, which were stuck on a double-sided adhesive conducting polymer tape. Surface charging effects were compensated by making use of the Kratos coaxial neutralization system. The binding energy scale was calibrated with respect to the Zr $3d_{5/2}$ component of the mixed oxide support, and fixed at 182.64 eV as reported in our previous work [16]. XPS data analysis was performed with CasaXPS Software, version 2.3.17dev6.3a, developed by Neal Fairley (Casa Software Ltd, 2013).

Transmission electron microscopy experiments were performed on two microscopes. A JEOL2010F field emission gun instrument equipped with an Oxford INCA Energy 2000 XEDS spectrometer was used to record High Angle Annular Dark Field (HAADF) images in the Scanning Transmission Electron Microscopy mode (STEM). STEM-HAADF images, whose intensity is roughly proportional to the square of the atomic number ($Z^2$), were obtained using an electron probe of 0.5 nm of diameter at a diffraction camera length of 8 cm. Based on the STEM-HAADF images of the catalysts, the diameters of more than 200 metal particles randomly selected were measured and the corresponding metal particle size distributions (PSD) were determined. Based on these PSDs, the average particle diameter (d) was calculated according to the following expression: d = $\sum n_i d_i / \sum n_i$, where $\sum n_i \geq 200$. Likewise, the total metal dispersion was calculated according to D = $N_s/N_t$, where $N_s$ is total number of surface metal atoms and $N_t$ is total number of atoms in the metal particle. For this last calculation, a home-made software (GAUSS) was used assuming a truncated cuboctahedron particle shape [16].

The compositional analysis of the nanoparticle systems present in each catalyst was performed by XEDS in STEM mode using a double aberration-corrected FEI Titan$^3$ Themis 60–300 microscope equipped with a high efficiency, high sensitivity, 4-detector ChemiStem system. Very high spatial resolution STEM-XEDS maps were acquired using a high brightness, sub-angstrom (0.07 nm) diameter, electron probe in combination with a highly stable stage which minimized sample drift. Element maps were acquired with a screen current of 80-120 pA and a pixel dwell time of 100-200 μs. This dwell time resulted in a frame acquisition time of approximately 20-30 s after which the residual drift was automatically corrected using cross correlation techniques. An averaging filter was used on the images as provided in the Esprit software from Bruker.

Every single particle included in collections containing more than one hundred of them was individually analyzed in each bimetallic catalyst. Quantification of the experimental STEM-XEDS provided compositional information from each analyzed particle. The Au:Pd molar ratio and Au content of each bimetallic catalyst was also calculated using the size and composition data of each metal particle analyzed by STEM-XEDS. Composition-size plots could also be established from this information.

*2.3. Catalytic activity for benzyl alcohol oxidation*

Catalytic tests were carried out in a thermostated glass reactor (30 mL) equipped with an electronically controlled magnetic stirrer connected to a large reservoir (5000 mL) containing oxygen [17]. The oxygen uptake was followed by a mass-flow controller. Benzyl alcohol and the catalyst (alcohol:total metal = 3000 mol:mol) were mixed in cyclohexane (benzyl alcohol:cyclohexane 50:50 vol:vol; total volume: 10 mL). The reactor was pressurized at 200 kPa of oxygen and heated to 80 °C. The reaction was initiated by stirring, periodic removal of samples from the reactor was performed. Identification and analysis of the products were done by comparison with the authentic samples by GC using a HP 7820A gas chromatograph equipped with a capillary column (HP-5 30 m 0.32 mm, 0.25 μm Film, made by Agilent Technologies) and thermal conductivity detector (TCD). Quantification of the reaction products was done by the external calibration method.

## 3. Results

### 3.1. Textural and structural properties

Table 1 gathers the BET specific surface areas of catalysts, which are very close to that of the ceria-zirconia mixed oxide support (67 $m^2g^{-1}$). It should be mentioned that after oxidation at 700 °C, the BET specific surface area of the 0.8AuPdCZO700 catalyst decreases slightly to 60 $m^2g^{-1}$.

XRD patterns of the catalysts indicate that there is no significant change of ceria-zirconia phase after oxidation at 450 or 700 °C (Figure 1). No Au, Pd or Au-Pd alloy diffraction peak was observed neither on the monometallic nor the bimetallic catalysts oxidized up to 450 °C. Therefore, Au, Pd or bimetallic Au-Pd metal particles are well dispersed on these catalysts. However, a new diffraction peak at 39°, i.e. between Au {111} and Pd {111} peaks, appears in the diagram of the 0.8AuPdCZO700 catalyst. The appearance of this diffraction peak suggests the formation of a Au-Pd alloy and increase of metal particle size after oxidation at 700 ºC [12,13].

*3.2. STEM results*

STEM-HAADF images of the monometallic Au/CZ and Pd/CZ catalysts were included in our previous article [12] and reproduced in Figure S1. Both of monometallic catalysts possess small average particle size of 1.7 and 1.9 nm and high metal dispersion of 39% and 45%, respectively. Figures 2, 3 and 4 present the results of the STEM analysis of the 0.8AuPdCZO250 catalyst pretreated at increasing temperatures. Particle size distributions of all the samples are shown in Figure 2. In addition, the elemental map distributions with low and high magnification are shown in Figures 3, 4 and S2.

In the catalyst treated at 250 ºC, most metal particles are smaller than 4 nm and their average particle size is 2.1 nm with metal dispersion of 36%. The STEM-XEDS analysis of nearly 120 individual metal particles clearly confirm the presence of three kinds of metal particles (Au, Pd and Au-Pd). This can be observed in the maps corresponding to this catalyst in Figures 3a, 3d and S2a. Note particularly, there are locations where only Au or Pd is detected whereas in other two catalysts both elements coincide.

The frequencies of each kind of metal particles, as well as the relationship between particle size and composition, are shown in Figure 2. More details, such as mean particle size of each kind of metal particle and average Au content both in the fraction of bimetallic particles and in the whole set of particles have been listed in Table S1.

According to the frequency plot, STEM analysis reveals that the Au-Pd bimetallic particles are dominant (71%), while monometallic Pd particles with diameters between 1 and 3 nm amount to 29% of the particles. The Au content in all the bimetallic particles fall within a very wide range of 10 to 80%. The average Au content of the whole set of particles estimated from STEM-XEDS is around 35%, which is a bit lower than that measured at macroscopic value (44%) by ICP.

To complement the nanoanalytical study of individual particles, XEDS analysis of two large areas of this sample was also performed as shown in Figure S2a. The average Au:Pd molar ratio obtained from two large areas is 0.4, which is similar to that obtained from the analysis of the 120 individual particles (0.54). This suggests that the statistical analysis of a big amount of particles and areas are representative. In any case it has to be taken into account that a small deviation in the Au or Pd percentages lead to large fluctuations in the Au:Pd molar ratio.

After oxidation at 450 °C, the range of particle size increases to 5.5 nm, and the average particle size increases also slightly from 2.1 to 3.0 nm (Figure 2e). At the same time, the metal dispersion remains very close (37%) to that of the original catalyst. From the elemental map images in Figures 3, 4 and S2b, it can be seen that Au and Pd distribute on this sample more homogeneously than on that treated at 250 °C. The analysis of roughly 100 individual nanoparticles indicates that the frequency of Au-Pd bimetallic particles significantly increases to 92%. In fact, Figures 3 and S2b show that the frequency of locations where only Pd is detected decreases. This result points out to a larger integration between the two noble metals after oxidation at 450 °C. Figure 2h demonstrates that the Au content range of individual bimetallic particles shrinks to 5-65% on this catalyst. The average Au content of this catalyst measured from XEDS is 39%.

Large areas of the 0.8AuPdCZO450 catalyst were also analyzed. Figure S2b shows the results of the analysis of 2 different areas and corresponding STEM images of this catalyst. The average Au content on these analyzed areas (32%) results in a Au:Pd molar ratio of 0.52.

The metal particle size spans the range from 2 to 15 nm in the 0.8AuPdCZO700 catalyst and the average particle size increases to 6.4 nm after oxidation at 700 ºC (Figure 2f). The metal dispersion of 0.8AuPdCZO700 drops to almost half (19%). The STEM results are in good agreement with XRD results, showing the appearance of Au-Pd alloy and increase of particle size. Oxidation at 700 ºC results in a nearly 100% of bimetallic particles. In fact, only a negligible fraction of monometallic particles could be detected by STEM-XEDS. Furthermore, Au and Pd are dispersed much more homogeneously than on both bimetallic catalysts oxidized at 250 and at 450 ºC. The Au content range in 75 analyzed particles, from 30 to 60%, is even narrower compared with the other two bimetallic catalysts. The average Au content in the whole set of particles measured by XEDS is 42%, quite similar to the ICP results (44%). The XEDS analysis of 3 areas of this sample (Figures 3c, 3f and S2c) indicates that Au and Pd distribute homogeneously, with an average Au content of 49%. This is also clearly observed in the elemental maps shown for this catalyst in Figures 3, 4 and S2c in which Au and Pd appear always together. Small monometallic Pd particles were rarely detected after oxidizing at 700 ºC.

Line-analysis of a large number of the bimetallic particles present in the three catalysts was also performed (Figures 4 and S3) in order to obtain additional information about the spatial distribution of Pd and Au. The bimetallic particles in the 0.8AuPdCZO250 catalyst depict an uneven and complicated Au and Pd distribution. As illustrated in Figure 4, bottom row profile 1, the first 2 nm in the path are Pd-rich, whereas in the remaining part Au dominates. It recalls, somehow, to a Janus-type of particle in which bimetallic parts rich in each one of the components are merged together. Oxidation at 450 ºC reveals a better integration of the two elements, with profiles, like that shown as profile 2 in Figure 4, in which Pd and Au appear more evenly distributed in space, with a Pd-rich surface and a Au-rich core structure. Additional line-analysis of this sample depicting the same features have been included in profiles 4-6 of Figure S3. Finally, Au and Pd distribute very homogeneously along the analyzed line on the bimetallic particle in the 0.8AuPdCZO700 catalyst, profile 3 in Figure 4 and profiles 7-9 in Figure S3, which clearly reveals a mixture at atomic level between the two metallic components after this treatment.

In summary, though the starting catalyst oxidized at 250 ºC is dominated by a fraction of bimetallic particles of widely differing composition, it also contains a fraction of pure monometallic Pd entities. Increasing the oxidation temperature to 450 or 700 ºC leads to a slight increase of the particle size, particularly at 700 ºC and an increase in the fraction of bimetallic Au-Pd nanoparticles and a homogenization in the composition of the nanoparticles, which change significantly in their nanostructural details.

*3.3. XPS results*

Table 2 summarizes the XPS results obtained on all the catalysts. Au:Zr and Pd:Zr molar ratios provide reliable information about relative amounts of Au and Pd atoms available on the surface of the catalysts. The analysis depths of Au and Pd of XPS have been calculated using data available from the literature [18,19]. For the Au 4f signal, the analysis depth falls between 5.4 and 5.8 nm, whereas that of Pd 3d reaches only between 4.6 and 5.0 nm. These values have to be used to interpret properly the observed Au:Zr and Pd:Zr ratios in the different catalysts. In principle, we would expect that in metallic particles with diameters below roughly 10 nm, all the metallic atoms involved would be within the range of the analysis. For larger particles, the atoms in the core could be excluded by XPS analysis. In any case, the fraction of atoms in these cores must represent a very small contribution, since the volume of these cores is negligible with respect to that of the outer parts.

The Au:Zr and Pd:Zr molar ratios decrease 24% and 30% after oxidation temperature at 450 ºC, respectively. As it can be observed in the STEM section, the diameter of most metal particles in the 0.8AuPdCZO450 catalyst are below 5.5 nm, which indicates that these metal particles are still in the range of analysis depth of Au and Pd by XPS. It seems therefore that the full quantitative analysis of the XPS signals in metal catalysts dispersed over the surface of nanocrystalline supports is more complex than that performed on flat surfaces of bulk materials. In any case, it is clear that the growth of particle size is affecting both Au:Zr and Pd:Zr molar ratios. The larger change in the later agrees with the better intermixing between Au and Pd in the catalyst treated at 450 ºC.

Finally, oxidation at 700 ºC leads to a dramatic decrease of Au:Zr and Pd:Zr molar ratios, which confirms the occurrence of sintering and homogenization between Au and Pd after oxidation at 700 ºC. Actually, the average particle size increases to 6.4 nm. It should be mentioned that after this treatment, only a small fraction of particles larger than 12 nm, whose cores would be out of range for XPS detection, appears. This would explain the much more significant drop in the Au:Zr and Pd:Zr ratios indicated in Table 2. Likewise, it is also interesting to note that the Au:Pd molar ratio remains around 0.4-0.5 in the catalysts oxidized in the range 250-450 ºC but increases to 0.6, after oxidation at 700 ºC. This suggests that Pd is more affected by nanostructural reorganization and sintering than Au, which is in good agreement with the detailed STEM-XEDS observations. These suggest a progressive dilution of Pd into Au, starting from a catalyst with a significant fraction of isolated monometallic Pd small nanoparticles, through a core-shell type of Au@Pd or AuPd@Pd particles, and ending in a fully alloy type situation.

The oxidation states of Au and Pd have also been deconvolved using XPS, as shown in Table 2 and Figure S4. There are two kinds of Au species ($Au^0$ and $Au^{\delta+}$) with binding energies around 84.1 and 85.3 eV on the surface of the 0.8AuPdCZO250 catalyst. After oxidation at higher temperatures $Au^{\delta+}$ species disappear and only metallic $Au^0$ species remain. In the case of Pd, two oxidation states are also found, but in contrast with Au, the percentage of $Pd^{\delta+}$ increases and the fraction of metallic $Pd^0$ decreases when the oxidation temperature is raised up to 450 ºC. Then, oxidation at 700 ºC does not induce any further change of Pd oxidation states in the bimetallic catalysts.

The fraction of $Ce^{3+}$ species does not change with oxidation temperatures. This confirms that the oxidation state of cerium is not a factor to explain changes in the catalytic activity for benzyl alcohol oxidation.

*3.4. Catalytic activities for selective oxidation of benzyl alcohol*

Figure 5 shows the benzyl alcohol conversion over all the catalysts. During 2 h of reaction time, Au/CZ catalyst is not active at all and the highest conversion of benzyl alcohol remains below 5%. The reference monometallic Pd/CZ catalyst is slightly more active than Au/CZ catalyst but the benzyl alcohol conversion only reaches 16.6%. No benzyl alcohol conversion was detected over the physical mixture of Au/CZ and Pd/CZ catalyst with the Au:Pd ratio of 0.8 during the first hour. However, it exhibits higher catalytic activity than both monometallic catalysts at reaction time of 2 h. This result is in good agreement with previous results on a physical mixture of Au and Pd supported on activated carbon (Au/AC and Pd/AC), which is also more active than the monometallic Au/AC and Pd/AC catalysts for the same reaction [2].

Likewise, a synergistic effect can be clearly observed over the bimetallic 0.8AuPdCZO250 catalyst, which exhibits the highest activity among all the catalysts. Oxidation at 450 ºC leads to a slight decrease of catalytic activity. Nevertheless, benzyl alcohol conversion declines dramatically for the 0.8AuPdCZO700 catalyst, for which the highest conversion reaches only 10% after 2 h of reaction time.

The selectivity values to benzaldehyde, benzyl benzoate and toluene are listed in Table 1 and S2. The main product over all the catalysts is benzaldehyde with selectivity above 95% during all the reaction time as shown in Table S2. The selectivity to benzyl benzoate is only 2-3% for all the bimetallic catalysts. Finally, a small amount of toluene (less than 3%) forms. Our previous article discussed the reaction pathways of selective oxidation of benzyl alcohol over monometallic and bimetallic catalysts [11]. Benzaldehyde and toluene are produced via an alcoxy intermediate, whereas benzyl benzoate forms via a carbonyloxyl intermediate with benzyl alcohol [11,20]. No benzene has been detected over any of the catalysts during 2 h of reaction time.

*3.5. Characterization of the used catalysts after reaction*

The 0.8AuPdCZO450, 0.8AuPdCZO700 and the physical mixture of Au/CZ and Pd/CZ catalysts after reaction have been studied by ICP, firstly (Table 3). The Au loadings decrease from 2.4% to 2.0% and 2.2% in the used 0.8AuPdCZO450 and 0.8AuPdCZO700 catalysts, respectively. Pd loading diminishes from 1.6% to 1.4% in both bimetallic catalysts. In molar terms, the effect of leaching on metal loading is more severe for Pd than for Au since Pd is much lighter than Au. Likewise, the impact of leaching decreases with increasing of oxidation temperature for Au, whereas it does not change significantly for Pd. Both elements seem to be mobilized under reaction conditions but Pd is more affected in terms of net atom removal from the surface of the catalysts. The Au and Pd concentrations after reaction are 0.8% and 0.7% for the physical mixture of Au/CZ+Pd/CZ catalyst. Compared with the Au and Pd contents before reaction, Au loss is up to 30%, while almost no Pd loss occurs. This result is similar to that observed in the 0.8AuPdCZO450 catalyst which has similar particle size.

Though an accurate comparison about the influence of leaching between the monometallic catalysts in the physical mixture and the bimetallic ones is complicated, it seems clear that in the bimetallic catalysts Pd loss is enhanced and that of Au slightly diminished. Related to this, the increased level of mixture between Au and Pd in the catalyst oxidized at 700 ºC may explain the observed decrease in Au leaching. Concerning Pd leaching, it should be taken into account that in the 0.8AuPdCZO450 catalyst Pd is mixed with Au, as revealed by STEM-XEDS results. The mixture degree with Au increases after oxidation at 700 ºC, but the size of the bimetallic particles also increases and, additionally, Pd moves into the bulk of these larger nanoparticles. A balance between these two factors could explain that the intensity of leaching for Pd remains roughly constant.

To study the nanostructural effects of leaching, a specific STEM-HAADF and STEM-XEDS analysis was performed both on the bimetallic and the physical mixture catalysts after reaction (Figures 6 and 7). For the catalysts oxidized at 450 and 700 ºC after reaction, the particle size distribution and average particle size are similar to those before reaction. However, the metal dispersion decreases from 42% to 17% and the average particle size increase to 6.9 nm for the physical mixture of Au/CZ and Pd/CZ after reaction. Figure 6 (right column) shows the Au and Pd maps on this physical mixture after benzyl alcohol oxidation. The elemental maps exhibit, in addition to monometallic particles, locations where both gold (Au-Lα map) and Pd (Pd-Lα map) are found together. Moreover, at these locations, the areas covered by Pd are larger than those covered by Au, which confirms the formation of a Au-core@Pd-shell nanostructure. This particular distribution of the two metallic elements is better observed in the intensity profiles recorded across individual nanoparticles, Figures 7 and S5 (right column). Note in particular how the signal of Pd peaks at positions at the edges of the particles whereas the Au signal peaks in the middle positions. Such distribution of the metallic elements suggests that Pd is migrated and deposited over the surface of the Au nanoparticles under reaction conditions.

This migration of Pd onto the Au particles seems linked to the synergy observed in the case of the physical mixture at reaction time of 2 h. Note, in fact, that no benzyl alcohol conversion has been observed during the first hour and then there is an activation period the following hour. This shape of the conversion curve vs reaction time over this physical mixture differs from that of the bimetallic catalysts. The Au-Pd interactions are already present from the very beginning of the reaction over these bimetallic catalysts. In particular, the occurrence of an activation process which may correspond to the slow, net, migration of Pd onto the surface of the Au cores to form bimetallic nanoparticles [2].

Left and middle columns in Figure 6 show the STEM-HAADF and Au and Pd maps of the used bimetallic catalysts. Large Au particles with diameters up to 24 nm were observed in the 0.8AuPdCZO450 after reaction, which are not detected in the catalyst before reaction. In addition, Pd particles with much smaller size than those present in the catalyst before reaction, dispersed both on the support and the surface of the Au particles are also observed.

High resolution STEM-HAADF and elemental maps (Figures 7a and 7d) indicate that particles with Au-core/Pd-shell or AuPd-core/Pd-shell structure are also present in the 0.8AuPdCZO450 catalyst after reaction. This means that not only Au agglomeration but also Pd migration onto Au occurs during the reaction. The average particle size of this catalyst increases from 3.0 to 3.7 nm, however due to the existence of large Au particles, the total metal dispersion drops dramatically from 37% to 10%. Migration of Pd atoms onto Au or Au-Pd nanoparticles to form Au@Pd or AuPd@Pd nanostructures might produce new active phases for selective oxidation of benzyl alcohol under reaction conditions. Let´s recall at this respect that Enache et al. pointed out the importance of core@shell structures in the performance of Au-Pd bimetallic catalysts [21].

The 0.8AuPdCZO700 catalyst after reaction still remains homogeneous bimetallic Au-Pd particles as shown in Figures 6b, 6e, 7b, 7e and S5b. No significant migration of Pd onto the original alloy type nanoparticles to form a significant fraction of core@shell particles or Au agglomeration have been observed over this catalyst. In fact, the average metal particle size and the dispersion remain the same as that of the catalyst before reaction. Moreover, Kiely et al. reported that homogeneous bimetallic Au-Pd particles supported on titania exihibit lower catalytic activity than the bimetallic Au-core/Pd-shell structure [8]. The low catalytic activity of 0.8AuPdCZO700 catalyst for selective oxidation of benzyl alcohol might be due to both the bigger particle size and the homogeneous mixture of Au and Pd alloy in this catalyst.

## 4. Discussion

Catalytic activity of the bimetallic Au-Pd systems for selective oxidation of benzyl alcohol depends on different parameters such as reaction conditions, total metal loadings, Au:Pd molar ratio, particle size distribution, particle composition and nanostructure and support, among others [2,3,5–11,22]. On the other hand, the particle size, composition and nanostructure of bimetallic Au-Pd catalysts are varied with the synthesis methods and pretreatment conditions. However, in the literature, the compositional homogeneity and nanostructure of metal particles based on statistical analysis of large amount of particles was not neither nor considered and performed [5].

In this work, we observed that oxidation at high temperatures induces the growth of metal particle size in parallel with an increase in compositional homogeneity of the Au-Pd metal particles. The large volume of the particle collections guarantees statistical significance of the compositional data. Moreover, the very good agreement found between macroscopic data (XPS, ICP) and those determined at the nanoscale by STEM-XEDS provide further evidence and reliability to the in-depth compositional analysis performed in this work. It is important to highlight that such a serious approach has never been reported in the literature related to Au:Pd bimetallic systems over ceria-zirconia support before and after reaction, to the best of our knowledge.

We have also observed that catalytic activity for oxidation of benzyl alcohol decreases with increasing oxidation temperatures. These results suggest that this reaction is sensitive to metal particle size. Our results confirm a synergy between the two metals in this reaction, nevertheless the STEM results also indicate the occurrence of Pd migration onto Au particles during catalytic reaction. Therefore, the synergy due to the interaction between the two metals, does not necessarily need to be embedded in the initial structure of the starting catalyst but it can be established under reaction conditions, leading to the formation of Au@Pd or AuPd@Pd core-shell nanoparticles which are very active for this reaction. This would explain why 0.8AuPdCZO250, which features a very heterogeneous composition of metal particles, comprising a significant fraction of very small Pd nanoparticles and medium sized Pd-rich Au-Pd alloy nanoparticles, shows the highest catalytic activity. Migration of the smallest Pd particles onto both Au and Au-Pd small particles would increase the extent of Au-Pd interaction over those initially existing in the medium sized ones. Moreover, STEM results showed that the DP method leads to a fraction of these particles being core@shell type in the catalyst before reaction.

In terms of intrinsic activity, TOF, decreases with increasing temperature, especially at 700 °C. This result suggests that not only composition and size plays an important role in the catalytic activity for benzyl alcohol oxidation but also that there is a large influence of the homogeneity of metal distribution. In this sense, a deleterious effect of increasing both homogeneity and size has been observed. This result differs from that obtained for glycerol oxidation, in which a bimetallic Au-Pd catalysts showed no significant influence of oxidation temperature on the intrinsic catalytic activity [12].

In the literature, the effect of calcination temperature on particle size and catalytic activity have been studied over bimetallic Au-Pd supported on MgO, activated carbon and $TiO_2$ catalysts [5,6,8,9]. The average particle size of bimetallic Au-Pd supported on MgO increased from 10 to 16.3 nm after calcination at 450 °C, while the conversion of benzyl alcohol decreased to around 24% [5]. For an activated carbon support, the median metal particle size increased from 5.0 nm to 20 nm after calcination of 400 °C, accompanying a decrease of 96% in TOF value [6]. The average metal particle size still maintained similar (from 4.6 to 5.2 nm) on a $TiO_2$ support, however the TOF value declined dramatically to half after calcination at 400 °C [6]. Therefore, compared with other supports [5,6,8], bimetallic Au-Pd supported on ceria-zirconia catalysts exhibit much better resistance against aggregation at high temperatures. The average metal particle size slightly increased from 2.1 to 3.0 nm maintaining the metal dispersion after oxidation at 450 ºC and the TOF value dropped only by 23 %. The oxidation at 700 ºC led to a bigger average particle size (6.4 nm), which is still much smaller than those on other supports in the literature and a decrease of TOF by 95 %, but similar results were observed on activated carbon support just after calcination at 400 ºC. Therefore, our results suggest that the occurrence of a strong interaction between the metal particles and the ceria-zirconia support stabilizes to a large extent bimetallic Au-Pd catalysts [6].

An important question to discuss refers to the impact of alloy formation on the intrinsic activity of the catalyst. To this end, it is interesting to compare the results of TOF values based on the metal dispersion estimated by STEM. These are the $TOF_{TEM}$ values in Table 1. A few facts become clear after this comparison. Firstly, Au is more active than Pd in benzyl alcohol oxidation, in terms of activity per surface atom. Secondly, at the beginning of reaction, the activity of the physical mixture corresponds roughly to that of the Au fraction. Thirdly, the bimetallic catalysts oxidized at the lower temperatures are one order of magnitude more active than pure Au, increasing oxidation temperature from 250 to 450 ºC having a small but significant (22% decay) effect on intrinsic activity. Importantly, the intrinsic activity of the metal atoms decreases steeply in the catalyst oxidized at the highest temperature, to reach the level of that of atoms in the monometallic Au catalyst.

The above-mentioned results certify that the formation of alloy-type nanoparticles, in which Au and Pd mix at atomic level, does not improve the activity of Au in the bimetallic catalysts for benzyl alcohol oxidation. Such increase seems to be related, instead, to the presence of particles in which Au-rich areas are in contact with Pd-rich ones, such as is the case of the nanoparticles present in the catalysts oxidized at 250 ºC or 450 ºC, especially the former. Such type of Au-Pd contacts may be present in the starting catalyst or may be produced due to mobilization by leaching of Pd and Au under reaction conditions.

In a recent paper, Carter et al. [10] related the activity in benzyl alcohol oxidation to an electronic perturbation of Au atoms at the surface of the bimetallic nanoparticles. Our results suggest that such perturbation cannot result from the mixture of Au and Pd at atomic level, since the TOF values, both the total ones and those determined by STEM, for the catalyst in which such situation prevails (0.8AuPdCZO700 catalyst) are quite similar to that detected in the monometallic catalyst. The Au:Pd ratio in this 0.8AuPdCZO700 catalyst is very close to that for which Carter et al. found the largest synergy (1:1).

XPS results indicate only a small increase in the percentage of $Au^0$ atoms in the bimetallic 0.8AuPdCZO250 catalysts with respect to the monometallic Au/CZ catalyst. Nevertheless, it must be considered that XPS is measuring the electronic states of mostly the whole collection of Au atoms, not only those at the surface of the nanoparticles. Such small difference does not properly justify the 25% drop in TOF values observed between 250 ºC and 450 ºC, especially when the dispersion remains exactly the same in both catalysts.

Therefore, the synergy between Au and Pd in bimetallic catalysts seems better related to electronic states, or cooperative effects, associated to contact zones between the two metals in Au@Pd and AuPd@Pd nanostructures. Henning et al. [23] have studied the influence of the Pd-shell thickness of the Au-core/Pd-shell nanocrystals on catalytic activity for selective oxidation of benzyl alcohol. The optimized Pd thickness was 2.2 nm with the core size of 7.1 nm over the unsupported Au-core/Pd-shell nanocrystals. However, in this work the Au-core/Pd-shell structure developed during the reaction, which rules out the possibility to establish the influence on this parameter in our bimetallic catalysts.

## 5. Conclusions

The influence of oxidation temperatures on the size, composition, nanostructure and catalytic performance for selective oxidation of benzyl alcohol of bimetallic Au-Pd supported on ceria-zirconia catalysts has been investigated in this work. In terms of activity, both a synergy between Au and Pd and a deactivation with increasing oxidation temperature have been confirmed.

On the basis of an unprecedented and in-depth characterization of the nanoparticle system present both in the monometallic, bimetallic and a physical mixture, which is considered the analysis at atomic scale of a large number of particles in each catalyst through STEM-based techniques, the changes in their nanostructure have been monitored and fruitfully related to changes in catalytic performance. Compared with other supports, the resistance against sintering at high temperatures of Au-Pd metal particles supported on ceria-zirconia is proven to be higher, which makes these catalysts less sensitive to thermal deactivation effects.

Characterization data before and after reaction reveal that the formation of a totally homogeneous Au-Pd alloy does not promote the intrinsic activity of Au. Instead, Au-core/Pd-shell structures seems to be at the roots of the synergy between the two metals on different types of supports, including the ceria-zirconia mixed oxide investigated in this work. Furthermore, a migration of Pd onto Au occurs during selective oxidation of benzyl alcohol reaction, which clearly contributes to enhance the creation of new Au-Pd contact areas, at which the reaction seems to proceed.

**Acknowledgement**

This work has been supported by MINECO/FEDER Program of the EU (ENE2017-82451-C3-2-R, MAT2016-81118-P and MAT2017-87579-R) and Junta de Andalucía (Project FQM-3994). J. J. Delgado thanks the MINECO/FEDER "Ramón y Cajal" Program.

**Table 1** Chemical-physical and catalytic properties of the monometallic catalysts and bimetallic catalysts oxidized at different temperatures

| Catalyst | Au loading (wt%)[a] | Pd loading (wt%)[a] | $S_{BET}$ ($m^2g^{-1}$) | Average particle size [b] (nm) | Metal dispersion[b] (%) | Au:Pd molar ratio [c] | $TOF_{TOTAL}$[d] ($h^{-1}$) | $TOF_{TEM}$[e] ($h^{-1}$) | Selectivity[f] (%) Benzaldehyde | Benzyl benzoate | Toluene |
|---|---|---|---|---|---|---|---|---|---|---|---|
| Au/CZ | 2.5 | - | 64 | 1.7 ± 0.1 | 39 | - | 162 | 415 | 100 | 0 | 0 |
| 0.8AuPdCZO250 | 2.4 | 1.6 | 65 | 2.1 ± 0.1 | 36 | 0.5:1 | 2690 | 7472 | 98 | 2 | <1 |
| 0.8AuPdCZO450 | 2.4 | 1.6 | 63 | 3.0 ± 0.1 | 37 | 0.6:1 | 2067 | 5586 | 96 | 3 | 1 |
| 0.8AuPdCZO700 | 2.4 | 1.6 | 60 | 6.4 ± 0.2 | 19 | 0.7:1 | 123 | 647 | 95 | 2 | 3 |
| Pd/CZ | - | 1.2 | 66 | 1.9 ± 0.1 | 45 | - | 75 | 167 | 100 | 0 | 0 |
| Physical mixture Au/CZ+Pd/CZ | 1.0 | 0.7 | - | - | 42 | 0.8 | 130 | 306 | 98 | 1 | 1 |

[a] The Au and Pd loadings were determined by ICP analysis
[b] Average particle size and metal dispersion were determined by STEM technique
[c] Au:Pd molar ratio obtained by STEM-XEDS
[d] $TOF_{TOTAL}$ values were calculated based on total amount of metals of the catalysts when the reaction time is 0.5 h
[e] $TOF_{TEM}$ values were calculated based on metal dispersion of the catalysts when the reaction time is 0.5 h
[f] Selectivity values were calculated when the reaction time is 2 h

Table 2 Quantified XPS data of the monometallic and bimetallic catalysts

| Catalyst | $Ce^{3+}$ in total amount of Ce (%) | Au $4f_{7/2}$ | | | | Pd $3d_{5/2}$ | | | | Molar ratios | | | |
|---|---|---|---|---|---|---|---|---|---|---|---|---|---|
| | | $Au^0$ (%) | Binding energy of $Au^0$ (eV) | $Au^{\delta+}$ (%) | Binding energy of $Au^{\delta+}$ (eV) | $Pd^0$ (%) | Binding energy of $Pd^0$ (eV) | $Pd^{\delta+}$ (%) | Binding energy of $Pd^{\delta+}$ (eV) | Zr:Ce | Au:Zr | Pd:Zr | Au:Pd |
| Au/CZ | 5 | 77 | 84.5 | 23 | 85.4 | - | - | - | - | 0.43 | 0.09 | - | - |
| 0.8AuPdCZO250 | 33 | 94 | 84.1 | 6 | 85.3 | 79 | 336.1 | 21 | 338.3 | 0.44 | 0.17 | 0.40 | 0.44 |
| 0.8AuPdCZO450 | 21 | 100 | 84.1 | 0 | - | 58 | 336.1 | 42 | 338.1 | 0.41 | 0.13 | 0.28 | 0.48 |
| 0.8AuPdCZO700 | 36 | 100 | 83.8 | 0 | - | 57 | 336.0 | 43 | 338.0 | 0.40 | 0.07 | 0.11 | 0.60 |
| Pd/CZ | 14 | - | - | - | - | 52 | 336.2 | 48 | 338.2 | 0.42 | - | 0.10 | - |

**Table 3** Chemical-physical and catalytic properties of the used catalysts after selective oxidation of benzyl alcohol

| Catalyst | Au loading (wt%)[a] | Pd loading (wt%)[a] | Average particle size [b] (nm) | Metal dispersion[b] (%) |
|---|---|---|---|---|
| 0.8AuPdCZO450 after reaction | 2.0 | 1.4 | 3.7± 0.4 | 10 |
| 0.8AuPdCZO700 after reaction | 2.2 | 1.4 | 6.5± 0.2 | 19 |
| Physical mixture Au/CZ+Pd/CZ after reaction | 0.8 | 0.7 | 6.9± 0.2 | 17 |

[a] The Au and Pd loadings were determined by ICP analysis

[b] Average particle size and metal dispersion were determined by STEM technique

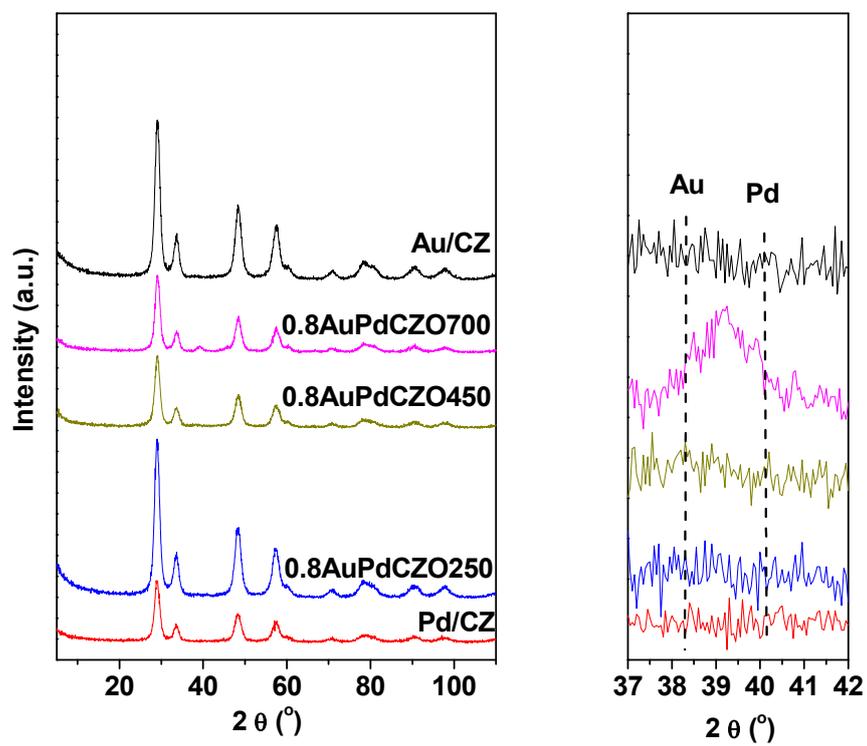

**Figure 1**. XRD patterns of the monometallic and bimetallic catalysts.

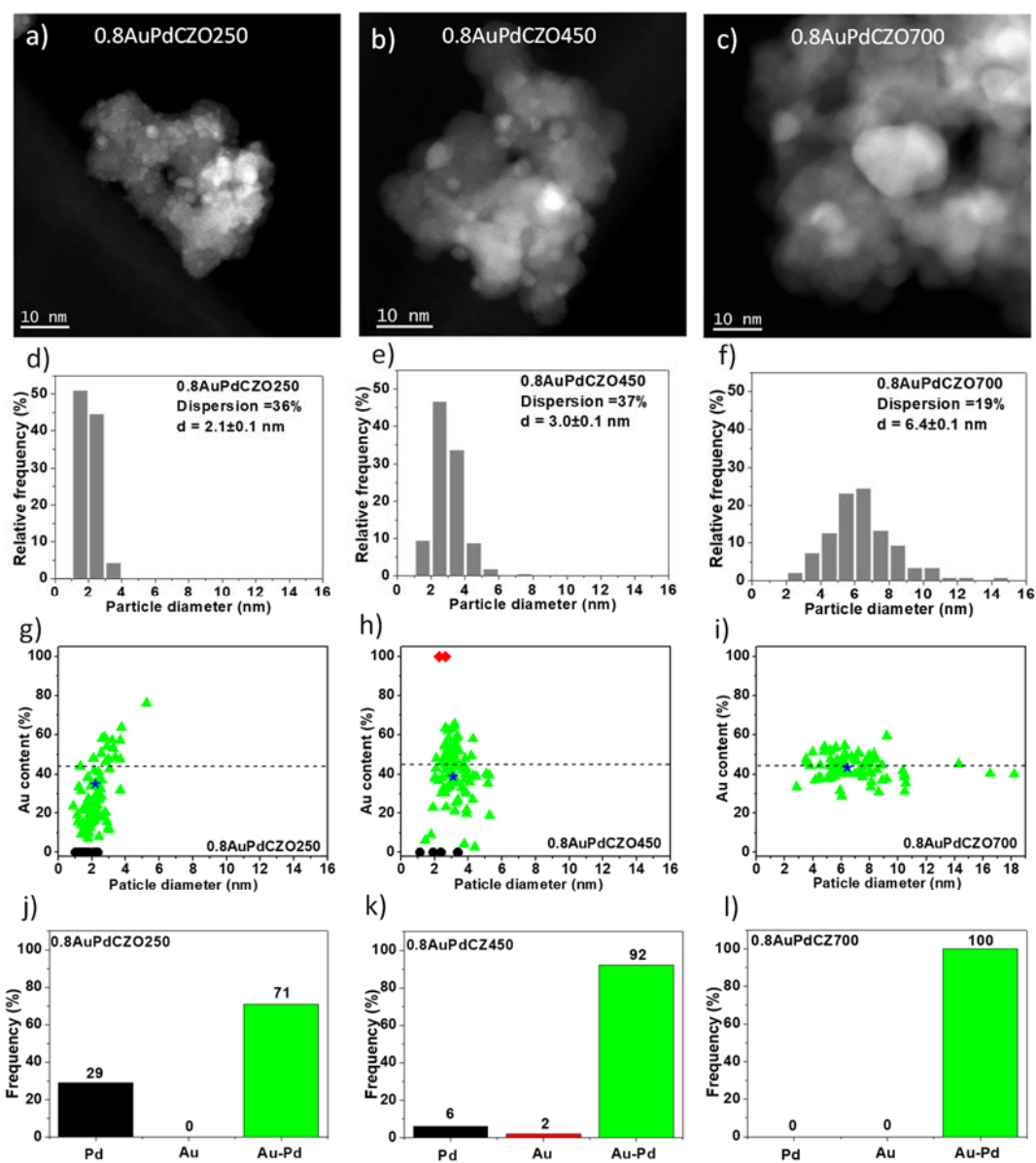

**Figure 2**. Representative STEM-HAADF images of 0.8AuPdCZ catalysts oxidized at different temperatures: (a) 250 °C, (b) 450 °C and (c) 700°C. The corresponding particle size distributions (d-f) determined by image analysis and the particle size versus composition diagrams (g-i) derived by STEM-XEDS of individual particles. The dashed line represents the Au:Pd composition obtained by ICP analysis. The blue star presents the average metal particle size and composition determined from the whole set of particles analyzed by STEM-XEDS. The black circles in the size vs composition plot correspond to monometallic Pd particles; green triangles to bimetallic nanoparticles and red diamonds to pure Au nanoparticles. (j-l) The frequencies of monometallic Pd and Au and bimetallic Au-Pd particles.

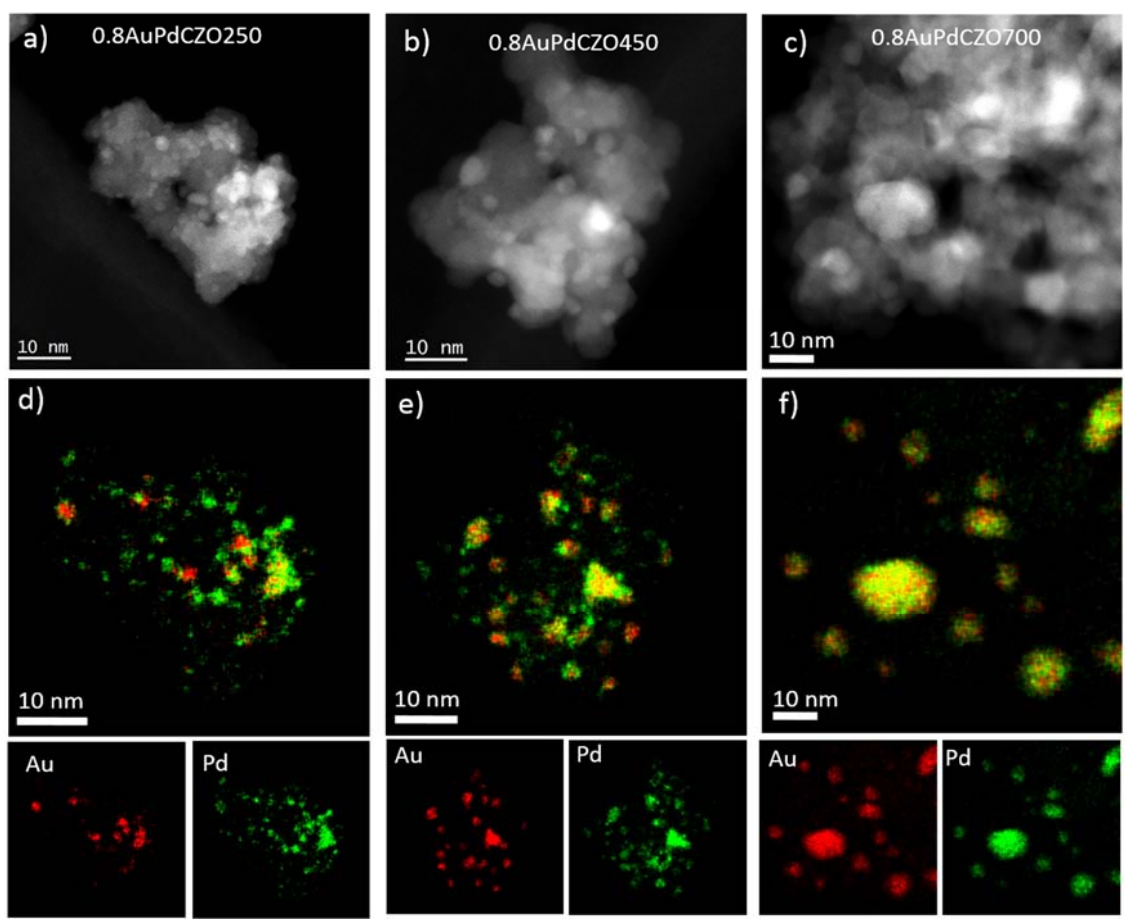

**Figure 3**. Middle magnification STEM-HAADF images and the corresponding XEDS elemental maps of Au-L (red) and Pd-L (green) for 0.8AuPdCZ catalysts oxidized at different temperatures ( a and d) 250 °C, (b and e) 450 °C, (c and f) 700 °C.

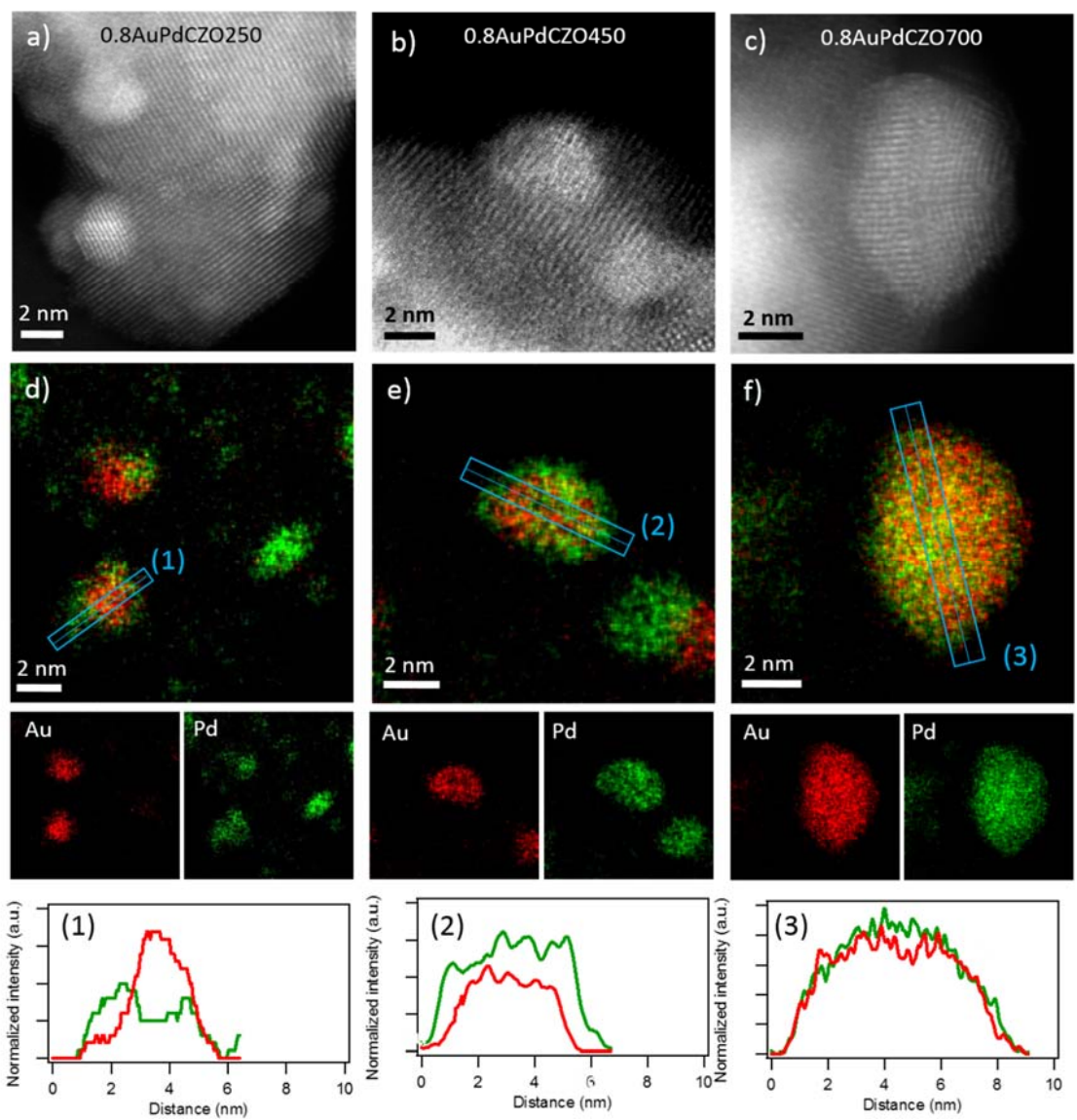

**Figure 4**. High magnification STEM-HAADF images (top) along the corresponding XEDS elemental maps of Au-L (red) and Pd-L (green) for particles of 0.8AuPdCZ catalysts oxidized at different temperatures: (a and d) 250 °C, (b and e) 450 °C and (c and f) 700 °C. At the bottom are represented XEDS Au-L (red) and Pd-L (green) linescan profiles taken across the diameter of each particle.

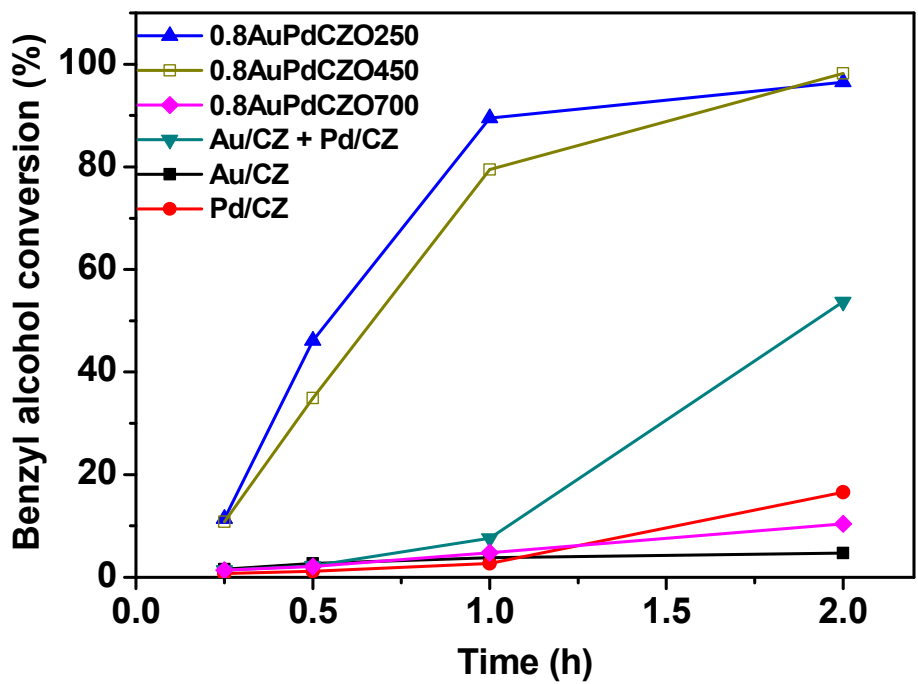

**Figure 5**. Catalytic activities of benzyl alcohol oxidation in liquid phase as a function of time in the presence of monometallic and bimetallic catalysts. Reaction conditions: alcohol:total metal ratio = 3000 mol:mol, benzyl alcohol:cyclohexane ratio = 50:50 vol:vol, reaction temperature: 80 °C. The reactor was pressurized at 200 kPa of oxygen.

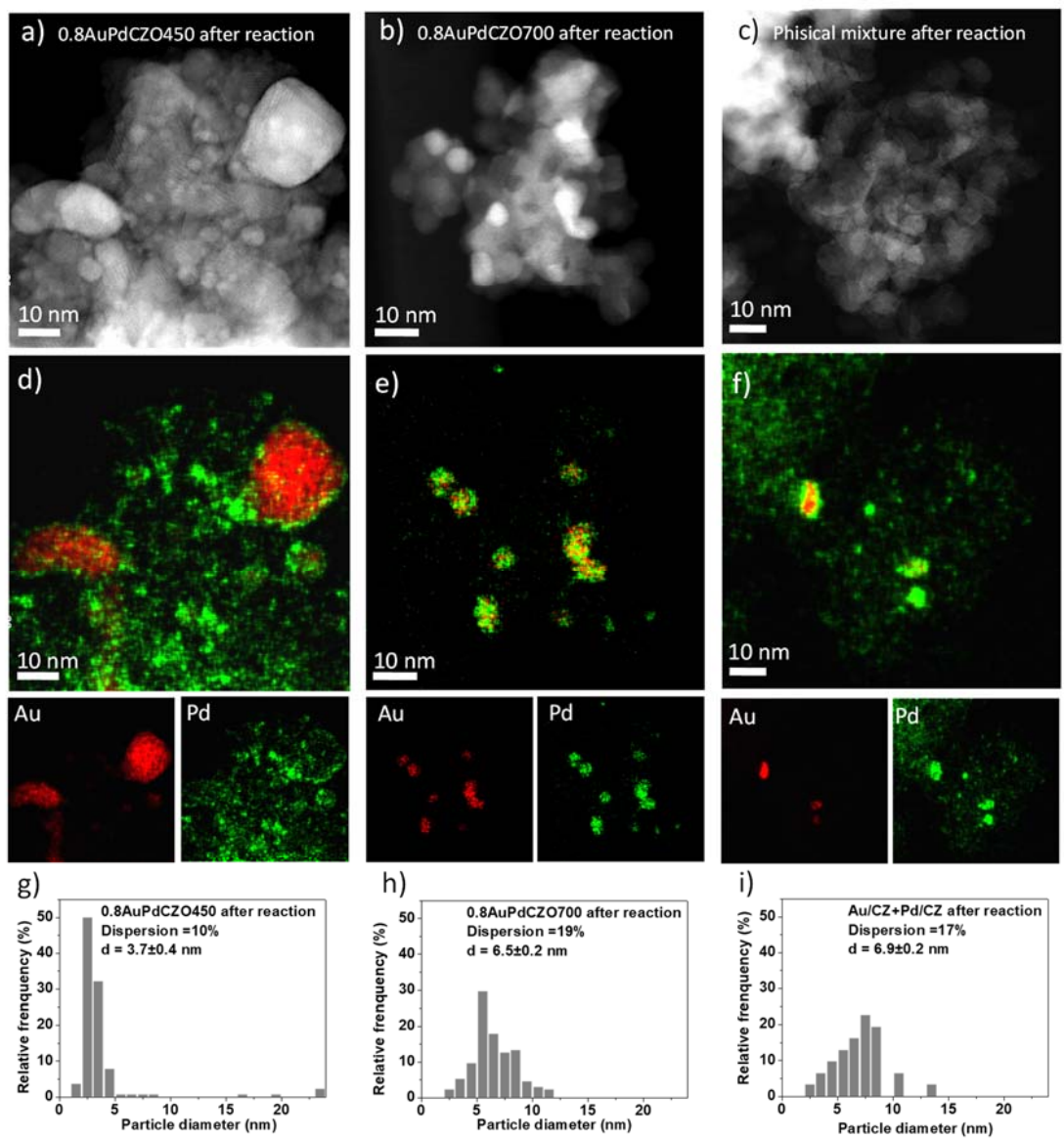

**Figure 6**. Representative middle magnification STEM-HAADF images (top) and XEDS elemental maps of Au-L (red) and Pd-L (green) of 0.8AuPdCZ catalysts after reaction: (a and d) oxidized 450 °C, (b and e) oxidized at 700 °C and (c and f) physical mixture of Au/CZ and Pd/CZ. Particle size distribution of the catalysts after reaction (g) 0.8AuPdCZO450, (h) 0.8AuPdCZO700 and (i) physical mixture of Au/CZ and Pd/CZ.

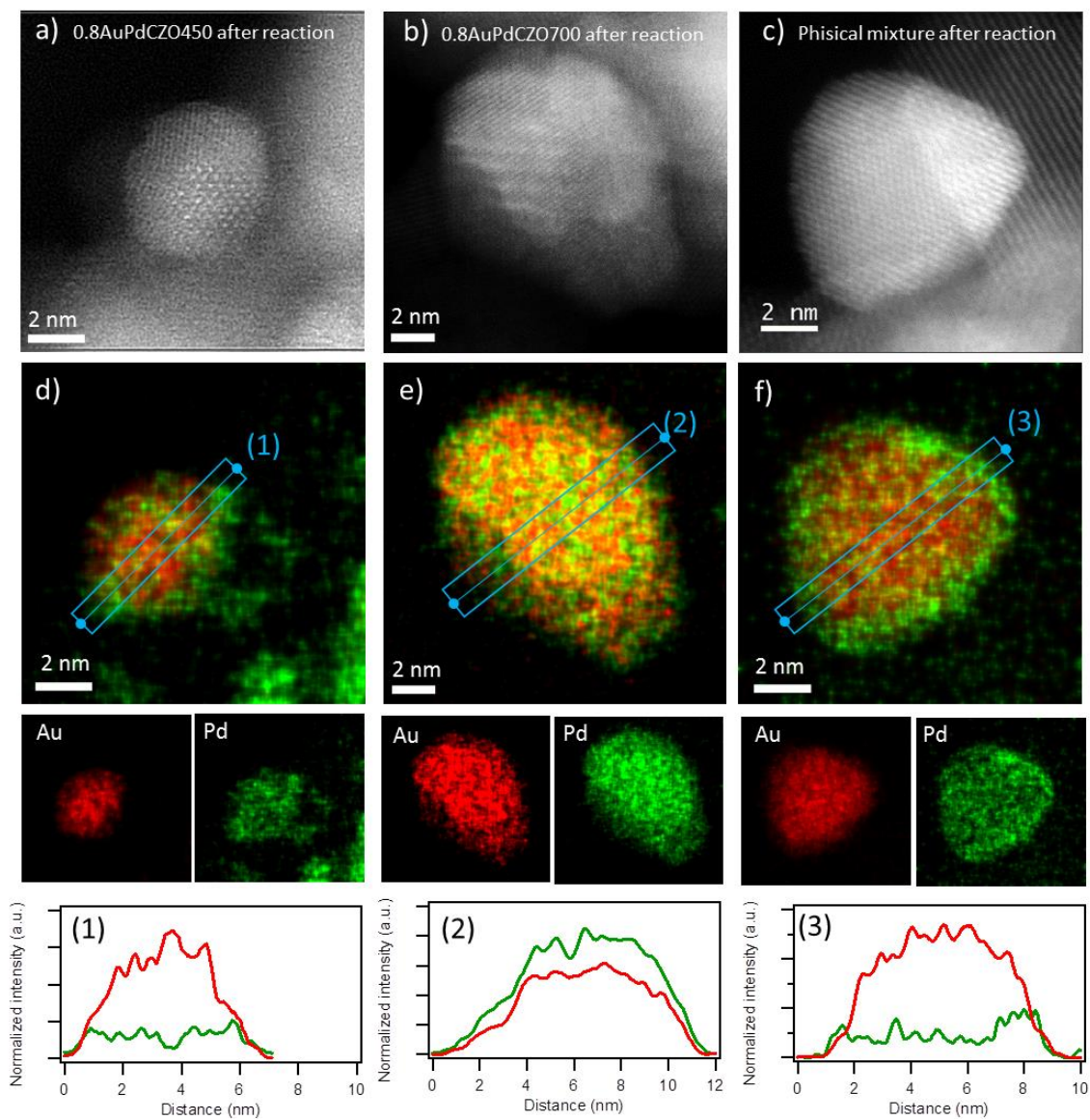

**Figure 7**. High magnification STEM-HAADF images (top) along the corresponding overlayed elemental maps of Au-L (red) and Pd-L (green) of bimetallic particles of 0.8AuPdCZ catalysts after reaction: (a and d) oxidized 450 °C, (b and e) oxidized at 700 °C and (c and f) physical mixture of Au/CZ and Pd/CZ. At the bottom are represented XEDS Au-L (red) and Pd-L (green) linescan profiles taken across the diameter of each particle.